\documentclass[twocolumn,prb,showpacs]{revtex4}
\usepackage{psfrag}
\usepackage{graphicx}
\usepackage{amsmath}
\usepackage{amssymb}

\begin{document}
\title{Partly occupied Wannier functions: Construction and applications}
\author{K. S. Thygesen}
\affiliation{Center for Atomic-scale Materials Physics, \\
Department of Physics, Technical University of Denmark, DK - 2800 Kgs. Lyngby, Denmark}
\author{L. B. Hansen}
\affiliation{Center for Atomic-scale Materials Physics, \\
Department of Physics, Technical University of Denmark, DK - 2800 Kgs. Lyngby, Denmark}
\author{K. W. Jacobsen}
\affiliation{Center for Atomic-scale Materials Physics, \\
Department of Physics, Technical University of Denmark, DK - 2800 Kgs. Lyngby, Denmark}

\date{\today}

\begin{abstract}
  We have developed a practical scheme to construct partly occupied,
  maximally localized Wannier functions (WFs) for a wide range of
  systems. We explain and demonstrate how the inclusion of selected
  unoccupied states in the definition of the WFs can improve both their
  localization and symmetry properties. A systematic selection
  of the relevant unoccupied states is achieved by minimizing the
  spread of the resulting WFs. The method is applied to a silicon
  cluster, a copper crystal and a Cu(100)
  surface with nitrogen adsorbed. In all cases we demonstrate the
  existence of a set of WFs with particularly good
  localization and symmetry properties, and we show that this set of
  WFs is characterized by a maximal average localization.
\end{abstract}
\maketitle

\begin{section}{Introduction}
\label{sec:intro}
A characteristic property of the single-particle eigenstates of most
molecular and solid state systems is their delocalized nature. For
many practical purposes this property is undesired and the construction of
equivalent representations in terms of localized orbitals becomes an
important issue.

Within the independent-particle approximation the use of Wannier
functions (WFs) allows for an exact description of the electronic
groundstate in terms of a minimal set of localized
orbitals\cite{wannier37}. The Wannier basis is truly minimal in the
sense that the number of orbitals is just enough to accomodate the
valence electrons of the system. Moreover, these localized WFs provide a formal justification of the widely used
tight-binding~\cite{ashcroft_mermin} and Hubbard models~\cite{mahan}.
Being the local analogue of the extended Bloch states of solid
state physics, the WFs formalize standard chemical concepts such as bonding,
coordination and electron lone pairs. Among the more technical applications of
Wannier functions we mention the connection to polarization
theory\cite{king-smith93,resta94} and their use within so-called
``linear scaling'' or ``order-$N$'' methods to obtain the electronic
groundstate\cite{goedecker99}. Very recently numerical methods for
electron transport calculations employing a Wannier function basis set have been
developed\cite{calzolari04,thygesen_chemphys}.

In the context of molecular systems the analogue of Wannier functions
for finite systems has been studied under the name "localized
molecular orbitals"~\cite{boys60, foster60, edmiston63, pipek89,berghold00,silvestrelli98}. These are traditionally defined by an
appropriate unitary transformation of the occupied single-particle
eigenstates and have been used for investigation of chemical
bonding. In the following we shall for simplicity use the
term WF to cover also localized molecular orbitals.

In 1997 Marzari and Vanderbilt developed a scheme to perform practical
calculations of maximally localized Wannier functions for an isolated
group of bands, i.e. a set of bands which is separated by a finite gap from all higher-
and lower-lying bands\cite{marzari_vanderbilt97}.  Within this scheme,
the usual arbitrariness inherit in the definition of the Wannier
functions due to the unspecified set of unitary transformations of
the Bloch states at every wave vector, is removed by requiring that
the sum of second moments of the resulting WFs is minimal. The method
follows the traditional idea of defining Wannier functions by a
unitary transformation of the occupied (Bloch-) orbitals. In general,
such methods fail to produce well localized orbitals when applied to
metallic systems because the unoccupied states
belonging to the partly filled valence bands\cite{iannuzzi02} are not
considered. Of
course, in cases where the partly filled valence bands are separated
by a gap from all higher bands, the method of Marzari and Vanderbilt
still applies. However, in the more general case where the bands of
interest cross and/or hybridize with other unwanted bands a different
approach must be used.

\begin{figure}[!b]
\includegraphics[width=0.85\linewidth,angle=0]{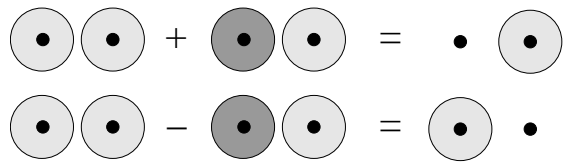}
\caption[cap.Pt6]{\label{fig.1sorbital} Schematic of the
  bonding-antibonding closure for a hydrogen molecule. The
  construction of well-localized atomic $s$-orbitals
  involves a matching of bonding and antibonding orbitals, independent
  of their occupation. The sign of the wave functions is indicated by
  the shading.}
\end{figure}
In this paper we demonstrate how the localization and in some cases
also the symmetry of a set of WFs can be drastically improved by
including selected unoccupied states in the definition of the
WFs~\cite{thygesen_WFprl}. The determination of the relevant
unoccupied states can be viewed as a bonding-antibonding closing
procedure, where occupied bonding states are paired with their
antibonding counterparts to yield localized orbitals. To be more
specific, consider two well-localized atomic orbitals on neighboring
atoms in a molecule. If we allow the two states to hybridize, a
bonding and an antibonding combination will result -- combinations
which may be less localized than the individual atomic orbitals.  To
regain the localized atomic orbitals from the molecular orbitals we
need both the bonding and the antibonding combination independent of
their occupation, see Fig.~\ref{fig.1sorbital}. In some cases the antibonding state may have
hybridized further with other states and the state which ``matches''
the bonding state will be a linear combination of eigenstates. The
problem we address here is the construction of a method for
systematically identifying the relevant unoccupied states. We show
that this can be achieved by optimizing the localization of the
resulting WFs. The paper gives a more detailed and extended account of
the work previously published in a Letter.\cite{thygesen_WFprl}

For periodic systems the bonding-antibonding
closure can be viewed as a procedure for disentangling the 
partly occupied valence bands from higher-lying bands. This problem
has previously been addressed by Souza \emph{et
  al.}~\cite{souza01} who proposed a disentangling method based on a
minimization of the change in character of the Bloch states across the 
 Brillouin zone (BZ). While this is a natural strategy for
crystalline systems, it is not clear how this
disentanglement procedure applies to non-periodic systems like
isolated molecules, a surface with adsorbates or a metal with impurities. 

The present method is related to that of Souza \emph{et
  al.}~\cite{souza01}, however, instead of minimizing the dispersion
across the BZ we suggest a disentanglement procedure based exclusively
on a minimization of the spread of the WFs. In this way we omit any
reference to the wave-vector and are therefore not limited to periodic
systems. The generality of the method is demonstrated by application
  to three different systems: an isolated $\text{Si}_5$ cluster, a
  copper crystal, and a Cu(100) surface with nitrogen adsorbed. Our results
  for the copper crystal are very similar to those obtained by Souza
  \emph{et al.}~\cite{souza01}, and this indicates the
similarity of the two localization schemes for periodic systems.

The paper is organized as follows:
In Sec.~\ref{sec.description} we introduce the spread functional and
outline the strategy behind the localization algorithm. In
Sec.~\ref{sec.gamma} we give the formal definition of partly occupied
WFs in the limiting case of a large supercell and derive the corresponding expressions for the gradient of the spread
functional. The
extension to periodic systems is discussed in Sec.~\ref{sec.kpoint}. In
Sec.~\ref{sec.results} we apply the method to a $\text{Si}_5$ cluster, a copper crystal and
a Cu(100) surface with adsorbed nitrogen.
\end{section}

\begin{section}{Description of the method}\label{sec.description}
In this section we introduce the spread functional used to
measure the degree of localization of a set of orbitals, and give an
introductory description of the localization scheme including its relation to the method of Souza \emph{et al.}~\cite{souza01}

\begin{subsection}{Spread functional}
\label{sec:spread_functional}
Within the localization scheme of Marzari
and Vanderbilt~\cite{marzari_vanderbilt97} the spread of a set of functions
$\{w_n(\bold r)\}_{n=1}^N$ is measured by the sum of second moments
\begin{equation}\label{eq.spreadfct1}
S=\sum_{n=1}^N (\langle w_n | r^2 | w_n \rangle - \langle w_n |\bold r
| w_n \rangle ^2).
\end{equation}
When periodic boundary conditions are applied, as in the present
study, and the supercell is sufficiently large, the minimization of $S$ is equivalent to
the maximization of~\cite{resta99}
\begin{equation}\label{eq.spreadfct2}
\Omega =\sum_{n=1}^N\sum_{\alpha=1}^{N_G} W_{\alpha} |Z_{\alpha,nn}|^2,
\end{equation}
where the matrix $Z_{\alpha}$ is defined as
\begin{equation}\label{eq.defx}
Z_{\alpha,nm}=\langle w_n|e^{-i\bold G_{\alpha} \cdot \bold r}|w_m \rangle.
\end{equation}
The $\{\bold G_{\alpha}\}$ is a set of at most six reciprocal lattice
vectors and $\{W_{\alpha}\}$ are corresponding weights
which account for the shape of the unit cell. For a definition and
discussion of these quantities we refer to
Refs.~\onlinecite{silvestrelli98,berghold00}.
\end{subsection}

\begin{subsection}{The localization scheme}
The starting point is the set of single-particle 
eigenstates, $\{\psi_n\}$, resulting from a conventional electronic
structure calculation. For simplicity we shall assume that the system
is isolated or is contained in a large supercell such that
reference to $\bold k$-points can be omitted. The aim is to obtain a
  set of $N_w$ localized WFs with the property that any eigenstate
  below a specified energy, $E_0$, can be exactly reproduced as a
  linear combination of the WFs. An obvious way to achieve this would
  be to apply the method of Marzari and Vanderbilt to compute the
  unitary transformation of the $N_w$ lowest eigenstates
  leading to the most localized WFs. The problem with
  this strategy is, however, that it is in general not possible to
  localize all WFs simultaneously, and the problem cannot be overcome
  by increasing $N_w$. 
 
  Instead, we define an external localization space as the space
  spanned by the $N_b$ lowest-lying eigenstates ($N_b>N_w$). Within
  this space we consider the subspace spanned by the eigenstates with
  energy below $E_0$, together with $L$ extra degrees
  of freedom (EDF). We shall refer to this subspace as the active
  localization space or simply the localization space. The EDF are
  assumed to be orthogonal and $L$ is chosen such that the dimension of the active localization space
  equals $N_w$. We then perform a simultaneous optimization of the
  WFs within the active localization space and of the active localization space
  itself. In practice this is achieved by optimizing an $N_w\times N_w$ unitary
  matrix together with the coordinates of the EDF such that the
  functional $\Omega$ becomes maximal. 

It is the determination of the EDF that distinguish our method from
that of Souza \emph{et al.}~\cite{souza01} In the latter, the spread functional is decomposed into
two terms: $\Omega=\Omega_I+\widetilde \Omega$, where $\Omega_I$
is related to the $k$-space dispersion of the band-projection
operator, see Ref.~\onlinecite{souza01}. In the first step, the
EDF are determined by maximizing $\Omega_I$, which depends only on the
localization space itself and not on the internal unitary
transformation. In the second step $\widetilde \Omega$, or
equivalently $\Omega$, is then maximized within the fixed localization
space. It is clear, that the separate maximization of $\Omega_I$ and
$\widetilde \Omega$ does not amount to the global maximization of
$\Omega$ that we propose here. We shall, however, see that the two methods lead to very
similar results in the case of periodic systems.
 
\end{subsection}
\end{section}

\begin{section}{Large supercells}
\label{sec.gamma}
In this section we give a detailed description of the localization scheme in the limiting
case of a large supercell where a $\Gamma$-point sampling of
the first Brillouin zone is a good approximation. For simplicity we
discuss this case separately before extending it to periodic
systems, although the latter contains the former as a special case. 
After giving the definition of 
partly occupied Wannier functions we derive expressions for
the gradients of the spread functional and discuss how to combine these
with a Lagrange multiplier scheme to determine the
maximum of $\Omega$.  

\begin{subsection}{Definition of partly occupied Wannier functions}
\label{sec:mathematicalstructure}
We denote the total number of eigenstates obtained from the electronic
structure calculation by $N_b$ and the number of eigenstates below the
energy $E_0$ by $M$. Our aim is to construct a set of $N_w$ WFs which
span at least the $M$ lowest-lying eigenstates. The remaining
$L=N_w-M$ degrees of freedom are simply used to improve the
localization of the resulting WFs as much as possible.  We expand
the WFs in terms of the $M$ lowest lying eigenstates and $L$ extra
degrees of freedom, $\{\phi_l\}$, belonging to the
$(N_b-M)$-dimensional space of eigenstates with energy above $E_0$:
\begin{equation}\label{eq.expansion1}
w_n=\sum_{m=1}^{M}U_{mn}\psi_m+\sum_{l=1}^{L}U_{M+l,n}\phi_l,
\end{equation}
where the extra degrees of freedom (EDF) are written as
\begin{equation}\label{eq.expansion2}
\phi_l=\sum_{m=1}^{N_b-M}c_{ml}\psi_{M+m}.
\end{equation}
The columns of the matrix ${c}$ are orthonormal and represent
the coordinates of the EDF with respect to the eigenstates lying above $E_0$. The matrix $U$ is unitary and represents a rotation of the
functions $\{\psi_1,\ldots,\psi_M,\phi_1,\ldots,\phi_L\}$.

In order to simplify the notation we introduce the matrices
\begin{equation}\label{eq.matrices}
{C}=
\left[ \begin{array}{cc} {I}^{M\times M} &  0 \\
 0 & {c}
\end{array}\right ]\; ,\quad
{V}=
{C}{U}=
\left[ \begin{array}{c} {U}^{M} \\
{c}{U}_{L}
\end{array}\right ],
\end{equation}
where $U^M$ and $U_L$ denotes the $M$ uppermost and
$L$ lowermost rows of $U$, respectively.
The $i$th column of $V$ gives the coordinates of
$w_i$ with respect to the full set of eigenstates $\{\psi_n\}$.

Substituting the expansions~(\ref{eq.expansion1}) and~(\ref{eq.expansion2}) into
Eq.~(\ref{eq.defx}) we obtain a compact matrix expression  
\begin{equation}\label{eq.xgeneral}
Z_{\alpha}=V^{\dagger} Z_{\alpha}^{(0)} V=
U^{\dagger}C^{\dagger} Z_{\alpha}^{(0)} C U,
\end{equation}
where $Z_{\alpha}^{(0)}$ is obtained from Eq.~(\ref{eq.defx}) by using the
eigenstates $\{\psi_n\}$ in the inner product,
\begin{equation}\label{eq.defx0}
Z^{(0)}_{\alpha,nm}=\langle \psi_n|e^{-i\bold G_{\alpha} \cdot \bold r}|\psi_m \rangle.
\end{equation}

\end{subsection}

\begin{subsection}{Gradient of $\Omega$}
\label{sec:gradient}
Through Eq.~(\ref{eq.xgeneral}) the spread functional, $\Omega$, in
Eq.~(\ref{eq.spreadfct2}) becomes a
function of the matrices $U$ and $c$. The maximum of $\Omega$ can be
found iteratively by updating $U$ and $c$ in the direction given by the
gradient. In the following we derive 
expressions for the gradient of $\Omega$.

We write the unitary matrix at iteration $n$ as
$U^{(n)}=U^{(n-1)}\exp(-A)$, where $A$ is an anti-hermitian matrix.
Since we are only concerned with small variations, we expand the
exponential to first order, i.e.  $\exp(-A)\backsimeq 1-A$. Inserting
this into Eqs.~(\ref{eq.xgeneral}) and~(\ref{eq.spreadfct2}) we find
\begin{equation}\label{eq.derivativeA}
\frac{\partial \Omega}{\partial
A_{ij}}=\sum_{\alpha=1}^{N_G} W_{\alpha}[Z_{\alpha,ji}(Z^{*}_{\alpha,jj}-Z_{\alpha,ii}^*)-Z_{\alpha,ij}^*(Z_{\alpha,ii}-Z_{\alpha,jj})]. 
\end{equation}
All matrices in this expression refer to iteration $n-1$. The new
rotation at iteration $n$ is then obtained by multiplying $U^{(n-1)}$
by $\exp[-d (\nabla_{A} \Omega)]$ where $d$ is the length of the
steepest-ascent step and $[\nabla_{A} \Omega]_{ij}=\partial
\Omega/\partial A_{ij}$.

We now turn to the problem of determining the steepest uphill
direction of $\Omega$ with respect to variations in $c$.
In general, for a real-valued function $f(z=x+iy)$ the direction of
steepest ascent with respect to $z$ is given by 
\begin{equation}\label{eq.steepestascend}
\frac{\partial f}{\partial z^*}\equiv
\frac{1}{2}(\frac{\partial f}{\partial x}+i\frac{\partial f}{\partial y}).
\end{equation}
To calculate the gradient $\partial \Omega/\partial c_{ij}^*$ we 
use that 
\begin{equation}\label{eq.derivative1}
\frac{\partial |Z_{\alpha,nn}|^2}{\partial c^*_{ij}}=Z_{\alpha,nn}\frac{\partial
  Z^{*}_{\alpha,nn}}{\partial c^{*}_{ij}}+Z^*_{\alpha,nn}\frac{\partial Z_{\alpha,nn}}{\partial
  c^*_{ij}}.
\end{equation}
From Eq.~(\ref{eq.xgeneral}) it follows that 
\begin{eqnarray}
\frac{\partial Z_{\alpha,nn}}{\partial
  c^*_{ij}}&=&\sum_{abcd}U^{\dagger}_{na}\frac{\partial
  C^{\dagger}_{ab}}{\partial c^*_{ij}}Z^{(0)}_{\alpha,bc}C_{cd}U_{dn}\\ \nonumber
&+&\sum_{abcd}U^{\dagger}_{na}
  C^{\dagger}_{ab}Z^{(0)}_{\alpha,bc}\frac{\partial C_{cd}}{\partial c^*_{ij}}U_{dn},
\end{eqnarray}
and from definition~(\ref{eq.matrices}) 
\begin{eqnarray}\label{eq.partielB}
\frac{\partial C_{nm}}{\partial c^*_{ij}}&=&0\\
\frac{\partial C^{\dagger}_{nm}}{\partial c^*_{ij}}&=&\delta_{m,M+i}\delta_{n,M+j}.
\end{eqnarray}
It is now easy to establish that 
\begin{eqnarray}\label{eq.partialXnn}
\frac{\partial Z_{\alpha,nn}}{\partial c^*_{ij}}=[Z^{(0)}_{\alpha} 
V]_{M+i,n}U^*_{M+j,n}\\ \label{eq.partialXnn2}
\frac{\partial Z^*_{\alpha,nn}}{\partial c^*_{ij}}=[(Z^{(0)}_{\alpha})^{\dagger} 
V]_{M+i,n}U^*_{M+j,n}.
\end{eqnarray}
Combining Eq.~(\ref{eq.derivative1}) with~(\ref{eq.partialXnn})
and~(\ref{eq.partialXnn2}) we arrive at the desired expression
\begin{equation}\label{eq.gradc}
\frac{\partial \Omega}{\partial c^*_{ij}}=\sum_{\alpha=1}^{N_G}W_{\alpha}[Z^{(0)}_{\alpha} V
\text{D}(Z^*_{\alpha})U^{\dagger}+ 
(Z^{(0)}_{\alpha})^{\dagger} V \text{D}(Z_{\alpha}) U^{\dagger}]_{M+i,M+j},
\end{equation}
where $\text{D}(Z_{\alpha})$ is a diagonal matrix with $(Z_{\alpha,nn})$ in
the diagonal.

To treat the constraint that the EDF $\{\phi_l\}$ should be orthonormal
during the maximization procedure we introduce the Lagrange
multipliers $\lambda_{ij}$ and perform an unconstrained maximization
of the functional
\begin{equation}\label{eq.omega_L}
\Omega_L=\Omega-\sum_{ij}\lambda_{ij}\langle \phi_i | \phi_j \rangle.
\end{equation} 
The Lagrange
multipliers are initially unknown and must be estimated at
each iteration. At the maximum we have $\nabla_{c^*} \Omega_L
=0$ which is equivalent to the condition
\begin{equation}
\nabla_{c^*}\Omega - c \mbox{
  $\lambda$}^{\text{T}}=0.
\end{equation}
Multiplying by $c^{\dagger}$ from the left leads to
\begin{equation}
\mbox{$\lambda$}^{\text T}=c^{\dagger}\nabla_{c^*} \Omega.
\end{equation}
This relation can be used to estimate the Lagrange multipliers at
each iteration. A step of length $d$ in the steepest uphill direction is thus
accomplished by adding to $c$ the matrix $d(1-c
c^{\dagger})\nabla_{c^*} \Omega$, followed by an orthonormalization of
the columns of $c$.  

\end{subsection}
\end{section}

\begin{section}{Periodic systems}
\label{sec.kpoint}
We consider a periodic system with a unit cell defined
by basis vectors $\bold a_1,\bold a_2,\bold a_3$ which in turn
define the basis vectors of the reciprocal lattice $\bold b_1,\bold b_2,\bold b_3$.
The Bloch states, $\{\psi_{n\bold k}\}$, resulting from the electronic
structure calculation are characterized by a band index
$n$ and a crystal momentum $\bold k$. The total number of bands is denoted
by $N_b$ and the number of eigenstates at a
given $\bold k$-point with energy below $E_0$
is denoted by $M_{\bold k}$. We assume a uniform sampling of
the first BZ such that any $\bold k$-point can be written as 
\begin{equation}\label{eq.kpoints}
\bold k=\frac{n_1}{N_1}\bold b_1 + \frac{n_2}{N_2}\bold b_2 +\frac{n_3}{N_3}\bold b_3 ,
\end{equation}
where $N_i$ is the number of $\bold k$-points in the direction $\bold
b_i$ and $n_i=0,\ldots,N_i-1$. Note that the $\Gamma$ point is always
included. With this convention the Bloch states,
$\{\psi_{n\bold k}\}$, correspond exactly to the $\Gamma$-point
eigenstates of the repeated cell defined by the extended basis vectors
$N_1\bold a_1,N_2\bold a_2,N_3\bold a_3$. An alternative way of
stating this correspondence is to say that the $\bold k$-points in
Eq.~(\ref{eq.kpoints}) fall on the reciprocal lattice of the repeated
cell, see Fig.~\ref{fig.kpoints}.
As we shall see below, this correspondence allows us to use the spread
functional $\Omega$ defined in Eq.~(\ref{eq.spreadfct2}) also for the
periodic system. We stress that the formalism developed in the
following section contains the $\Gamma$-point formalism described in the
preceding sections as a special case.

\begin{subsection}{Definition of partly occupied Wannier functions}
We write the $n$th Wannier function related to unit cell $i$ as
\begin{equation}\label{eq.periodicWF}
w_{i,n}=\frac{1}{\sqrt{N_k}}\sum_{\bold k} e^{-i\bold k \cdot \bold
  R_i}\tilde \psi_{n\bold k},
\end{equation}
where $N_k$ is the total number of $\bold k$-points and $\tilde
\psi_{n\bold k}$ is a generalized Bloch state to be defined
below~\cite{marzari_vanderbilt97}. Each generalized band, i.e. each
set $\{\tilde \psi_{n\bold k}\}$ for fixed $n$, gives rise to one WF
per unit cell. These WFs are simply related by translation, i.e. $w_{i,n}(\bold
r)=w_{0,n}(\bold r-\bold R_i)$, and thus it suffices to consider the 
WFs of the cell at the origin. In doing this we can
omit the cell index and simply denote the WFs by $\{w_n\}$.
We denote the number of WFs per cell by $N_w$.

Following the idea behind Eq.~(\ref{eq.expansion1}) we
expand the generalized Bloch state $\tilde \psi_{n\bold k}$ in terms
of the $M_{\bold k}$ lowest lying Bloch states and $L_{\bold k}$ extra
degrees of freedom,
$\{\phi_{l\bold k}\}$, from the remaining $(N_b-M_{\bold
  k})$-dimensional space
\begin{equation}\label{eq.periodicexpansion1}
\tilde \psi_{n \bold k}=\sum_{m=1}^{M_{\bold k}}U^{\bold k}_{mn}\psi_{m\bold
  k}+\sum_{l=1}^{L_{\bold k}}U^{\bold k}_{M_{\bold k}+l,n}\phi_{l\bold k},
\end{equation}
where the EDF are expanded as
\begin{equation}\label{eq.periodicexpansion2}
\phi_{l\bold k}=\sum_{m=1}^{N_b-M_{\bold k}}c^{\bold
k}_{ml}\psi_{M_{\bold k}+m,\bold k}.
\end{equation}
The number of EDF at a given $\bold k$-point is determined by the condition $L_{\bold k}+M_{\bold
  k}=N_w$. If $M_{\bold
  k}$ exceeds $N_w$, we simply put $M_{\bold
  k}=N_w$.
Due to the exact correspondence between the Bloch states
$\{\psi_{n\bold k}\}$ and the $\Gamma$-point
eigenstates of the repeated cell, we can use the functional
(\ref{eq.spreadfct2}) to measure the spread of the Wannier
functions. The matrices $Z_{\alpha}$ are still defined by Eq.~(\ref{eq.defx})
but it should be remembered that the inner product as well as the
reciprocal lattice vector $\bold G_{\alpha}$ now refer to the repeated cell.
From Eqs.~(\ref{eq.periodicexpansion1},\ref{eq.periodicexpansion2}) we
find the following generalization of Eq.~(\ref{eq.xgeneral}) 
\begin{equation}\label{eq.xgeneralperiodic}
Z_{\alpha}=\sum_{\bold k,\bold k'}Z_{\alpha}^{\bold k \bold k'},
\end{equation}
where
\begin{equation}
Z_{\alpha}^{\bold k \bold k'}=(U^{\bold k})^{\dagger}(C^{\bold k})^{\dagger} Z_{\alpha}^{(0),\bold k\bold k'}
  C^{\bold k'} U^{\bold k'}.
\end{equation}
The matrix $C^{\bold k}$ is given by the obvious $\bold k$-point
analogue of Eq.~(\ref{eq.matrices}) and the matrix $Z_{\alpha}^{(0),\bold
  k\bold k'}$ is defined by
\begin{equation}
Z_{\alpha,nm}^{(0),\bold k\bold k'}=\langle \psi_{n\bold k}|e^{-i\bold
  G_{\alpha} \cdot \bold r}|\psi_{m\bold k'} \rangle.
\end{equation}
Most of the matrices $Z_{\alpha}^{(0),\bold k\bold
  k'}$ are in fact zero. Writing the Bloch functions as $\psi_{n\bold
  k}=u_{n\bold k}(\bold r)\exp(i\bold k\cdot \bold r)$, where
  $u_{n\bold k}$ has the periodicity of the lattice, we get
\begin{equation}
Z_{\alpha,nm}^{(0),\bold k\bold k'}=\int u^*_{n\bold k}(\bold r)u_{m\bold
  k'}(\bold r)e^{i(\bold k'-\bold k-\bold G_{\alpha})\cdot \bold r}\text{d}\bold r,
\end{equation}
which is non-zero only when
\begin{equation}\label{eq.condition}
\bold k'=\bold k+\bold G_{\alpha}.
\end{equation}
Here it is implicit that $\bold k$ and $\bold k'$ belong to the first BZ
and thus it might be necessary to translate $\bold k'$ by a reciprocal
lattice vector. 
The relation between 
$\bold k$ and $\bold k'$ is illustrated in Fig.~\ref{fig.kpoints}.
Note that the condition in Eq.~(\ref{eq.condition}) reduces the double sum
in Eq.~(\ref{eq.xgeneralperiodic}) to a single sum over $\bold k$.

\begin{figure}[!h]
\includegraphics[width=0.65\linewidth,angle=0]{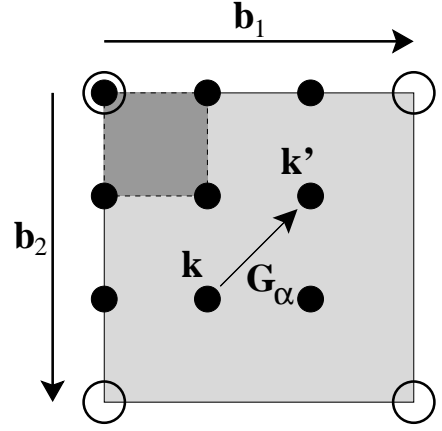}
\caption[cap.Pt6]{\label{fig.kpoints}
Relation between the first BZ of the unit cell,
defined by the reciprocal basis vectors $\bold b_1,\bold b_2,\bold b_3$
(light gray), and the first BZ of the repeated unit cell (dark gray).
In this case $N_1$ and $N_2$ from Eq.~(\ref{eq.kpoints}) both equals 3.
The relation between $\bold k$ and $\bold k'$, given in Eq.~(\ref{eq.condition}),
is indicated.}
\end{figure}

The derivation of the gradient of $\Omega$ follows closely the
$\Gamma$-point case discussed in Sec.~\ref{sec:gradient} and is
therefore omitted. The result is
\begin{widetext}
\begin{equation}
\frac{\partial \Omega}{\partial
A^{\bold k}_{ij}}=\sum_{\alpha=1}^{N_G}
W_{\alpha}[(Z_{\alpha,jj})^*Z_{\alpha,ji}^{\bold k-\bold G_{\alpha}, \bold
  k}+Z_{\alpha,jj}(Z_{\alpha,ij}^{\bold k, \bold
  k+\bold G_{\alpha}})^*-(Z_{\alpha,ii})^*Z_{\alpha,ji}^{\bold k, \bold
  k+\bold G_{\alpha}}-Z_{\alpha,ii}(Z_{\alpha,ij}^{\bold k-\bold G_{\alpha}, \bold k})^*]. 
\end{equation}
\begin{equation}
\frac{\partial \Omega}{\partial
  (c^{\bold k}_{ij})^*}=\sum_{\alpha=1}^{N_G}W_{\alpha}[Z_{\alpha}^{(0),\bold
  k,\bold k+\bold G_{\alpha}} V^{\bold k+\bold G_{\alpha}}
\text{D}(Z^*_{\alpha})(U^{\bold k})^{\dagger}+ 
(Z_{\alpha}^{(0),\bold k-\bold G_{\alpha},\bold k})^{\dagger} V^{\bold
  k-\bold G_{\alpha}} \text{D}(Z_{\alpha})
  (U^{\bold k})^{\dagger}]_{M_{\bold k}+i,M_{\bold k}+j}.
\end{equation}
\end{widetext}
We note that these expressions, of course, reduce to
Eqs.~(\ref{eq.derivativeA},\ref{eq.gradc}) in the limit of a single
$\bold k$-point. The maximization of $\Omega$ proceeds along the same lines as
for the $\Gamma$-point case, except that Lagrange multipliers are
needed for each
$\bold k$-point. For example the analogue of Eq.~(\ref{eq.omega_L}) reads 
\begin{equation}
\Omega_L=\Omega-\sum_{ij,\bold k}\lambda_{ij,\bold k}\langle
\phi_{i\bold k} | \phi_{j\bold k} \rangle.
\end{equation}  

\end{subsection}

\begin{subsection}{Optimizing the number of extra degrees of
    freedom}\label{sec.optimal} 
For given values of $N_b$, $N_w$ and $E_0$, the algorithm introduced
above produces the $N_w$ most localized WFs that can be formed within
the external localization space when all eigenstates below $E_0$
should be exactly reproducible in terms of the WFs. It remains to
determine the optimal values for $N_b$ and $N_w$ for a given
$E_0$. Let us start by considering the situation where $N_b$ has been
fixed at a value which is large enough to include all anti-bonding
states relevant for the localization. In practice this typically
means $\sim 10$~eV~above the Fermi level.  
It seems as a natural strategy to choose $N_w$ such that the localization
\emph{per} orbital is maximal. To quantify this condition we define the average
localization per orbital as
\begin{equation}\label{eq.averagespread}
\langle \Omega \rangle=\frac{\Omega[E_0,N_b,N_w]}{N_w},
\end{equation}
where we have indicated the dependence of $\Omega$ on the three
parameters explicitly. We note that since the value of $\Omega$ also
depends on the size and shape of the supercell, it does not
make sense to compare the value of $\Omega$ for systems described in
different supercells. Fixing $N_w$ on the basis of $\langle \Omega
\rangle$ represents a completely general criterion which can be applied in
any situation. However, the localization procedure must be carried out
for several values of $N_w$ which might be a tedious task depending on the size of
the system. We next consider the situation when $N_b$ is also allowed
to change. Formally, the global maximum of $\langle \Omega \rangle$ 
is attained in the limit where both $N_b$ and $N_w$ tend to infinity in
which case an infinite set of completely localized delta functions can be realized.
However, we have found that for practical values of $N_b$ where very
high energy states are not included in the external localization
space, $\langle \Omega \rangle$ will have a local maximum for some
$N_w$, and the position of the maximum is not sensitive to the actual value of $N_b$. 
Thus, it is indeed possible to determine an optimal value of $N_w$ by
maximizing $\langle \Omega \rangle$. 

Alternatively it is often possible to determine a value for $N_w$ based
on symmetry arguments, chemical intuition, or a closed band condition.
As we shall see in the following
examples the two criteria for determining $N_w$ lead to similar results.
\end{subsection}

\begin{subsection}{Start guess for $U^{\bold k}$ and $c^{\bold k}$}
  For small systems we have found that the
  localization algorithm is quite stable and usually leads to the global maximum
  independently of the initial value of the matrices $\{U^{\bold k}\},\{c^{\bold
  k}\}$. For larger systems, however, there is a risk of getting stuck
  in a local maximum and in such cases the start guess becomes
  important. It is then natural to start from a set of simple
  orbitals located either at the atoms or at the bond centers. Let
  $\{f_{\nu}\}$ denote such a set of simple orbitals. The question is
  how to transform this into the matrices $\{U^{\bold k}\},\{c^{\bold k}\}$.  To
  this end we project the initial orbitals onto the subspace spanned
  by the Bloch states at each $\bold k$-point:
\begin{equation}
\tilde f_{\nu \bold k}=\sum_{n=1}^{N_b} \langle \psi_{n\bold k}|f_{\nu}\rangle\psi_{n\bold k}.
\end{equation}

The following procedure is carried out for each $\bold k$-point
separately. For fixed $\bold k$ we regard $\langle \psi_{n\bold
  k}|f_{\nu}\rangle$ as a matrix in the indices $n,\nu$. Its
columns represent the coordinates of the $\tilde f_{\bold k
  \nu}$ with respect to the Bloch states $\{\psi_{n\bold
  k}\}_{n=1}^{N_b}$ and as such it is a (non-orthogonalized) version of
the matrix $V^{\bold k}$, see Eq.~(\ref{eq.matrices}).
After a normalization
of the columns of $\langle \psi_{n\bold k}|f_{\nu}\rangle$ we compute the norm of the
component of $\tilde f_{\nu \bold k}$ orthogonal to the occupied subspace:
\begin{equation}
\|\tilde f_{\nu \bold k}^{\perp}\|^2=\sum_{n=M(\bold k)}^{N_b}|\langle \psi_{n\bold k}|f_{\nu}\rangle|^2.
\end{equation}
The first EDF is chosen as a normalized version of the $\tilde
f_{\bold k\nu}^{\perp}$ for which
$\|\tilde f_{\bold k\nu}^{\perp}\|$ is the largest. The remaining
$\tilde f_{\bold \nu}^{\perp}$'s
are then orthogonalized onto this vector and the process is repeated
until all EDF, and thus $c^{\bold k}$, have been determined. Finally the identity 
$U^{\bold k}=(C^{\bold k})^{\dagger}V^{\bold k}$ with $V^{\bold k}\to \langle \psi_{n\bold
  k}|f_{\nu}\rangle$ determines $U^{\bold k}$. Since
the $\tilde f_{\nu \bold k}$ are not necessarily orthogonal, the
columns of the resulting $U^{\bold k}$ must be explicitly
orthogonalized.
\end{subsection}
\end{section}

\begin{section}{Results}\label{sec.results}
  In the following sections we apply the localization scheme to three
  different systems. To demonstrate the generality of the method we
  consider both isolated and metallic systems as well as a metal
  surface with adsorbed impurities. In Sec.~\ref{sec.si5} we construct
  partly occupied WFs for an isolated $\text{Si}_5$ cluster and
  illustrate how different sets of WFs can be obtained by varying the
  number of extra degrees of freedom. In Sec.~\ref{sec.cufcc} we
  investigate the WFs of a Cu(fcc) crystal and compare the results
  with those obtained by Souza and co-workers~\cite{souza01} who
  studied the same system using a different but related method.
  Finally, in Sec.~\ref{sec.ncu} we perform a detailed WF analysis for
  a Cu(100) surface with 0.5 mono-layers of nitrogen. In all
  calculations we use a plane-wave based DFT code~\cite{dacapo} to
  obtain the Kohn-Sham eigenstates, and we describe the ion potential
  by Vanderbilt ultrasoft pseudopotentials~\cite{vanderbilt90}. To
  ensure a proper convergence of the unoccupied states employed in the
  localization scheme, the DFT calculations have been converged with
  respect to the full set of Kohn-Sham eigenvalues. In
  the appendix we explain how to extend the localization scheme to
  ultrasoft pseudopotentials.

\begin{subsection}{$\text{Si}_5$ cluster}\label{sec.si5}
  As an example of an isolated system we consider an
  $\text{Si}_5$ cluster in its ground-state
  geometry~\cite{raghavachari85}, see Fig.~\ref{fig.si5_orbitals}(a).
  We use a cubic supercell of length 16~\AA~and sample the first BZ at the
  $\Gamma$-point. To test the dependence on the size of
  the external localization space we consider the two cases $N_b=30$
  and $N_b=100$. We set $M=10$ corresponding to the number of occupied
  states, and calculate the average localization per WF, $\langle
  \Omega \rangle=\Omega/N_w$, for $L=0,\ldots,7$. The result is shown
  in Fig.~\ref{fig.si5_spread}.  For $L=0$ there is no difference
  between the two cases since the WFs are constructed entirely from
  the occupied eigenstates. However, for $L\geq0$ the larger space
  available for the extra degrees of freedom leads to an improved
  localization when $N_b=100$. Apart from this general improvement in
  localization, there is no qualitative difference between the WFs
  obtained with $N_b=30$ and $N_b=100$ for a given $L$. We note that
  both curves have a maximum for $L=4$, corresponding to a total of 14
  WFs. This particular set of WFs together with their centers is shown
  in Fig.~\ref{fig.si5_orbitals}(b). The fact that this set of WFs
  respect the symmetry of the cluster is a special property of the
  $L=4$ solution: for other values of $L$, including $L=0$, the WFs
  break the symmetry of the $\text{Si}_5$ cluster. This indicates that
  the solution corresponding to the maximal value of $\langle \Omega
  \rangle$ has a special meaning. Indeed, the value $N_w=14$ could
  also have been anticipated from physical arguments. Starting from a
  set of four $sp^3$ orbitals located at each Si atom we expect
  bonding and anti-bonding states to form between pairs of aligned
  orbitals belonging to nearest neighbor pairs of Si atoms. 
  These bonding states can be identified as the six bond-centered WFs
  shown to the far left in Fig.~\ref{fig.si5_orbitals}. The two ``top'' Si
  atoms have three nearest neighbors and thus a single $sp^3$ orbital
  is left as a lone pair (middle WF). The remaining three Si atoms
  each have two nearest neighbors and consequently two $sp^3$ orbitals
  are left as lone pairs (rightmost WF). In total this adds up to
  14 orbitals. The anti-bonding counterparts of the bonding states
  formed between nearest neighbors are not brought into play for
  $L=4$, because they are much less localized than the bonding states.
  However, by setting $L=10$ and thus searching for a total of 20 WFs,
  the anti-bonding states are picked out as EDF and we obtain a full
  set of $sp^3$ orbitals. This solution has, however, a smaller value
  for $\langle \Omega \rangle$ than the solution at $L=4$.
\begin{figure}[!h]
\includegraphics[width=1.0\linewidth]{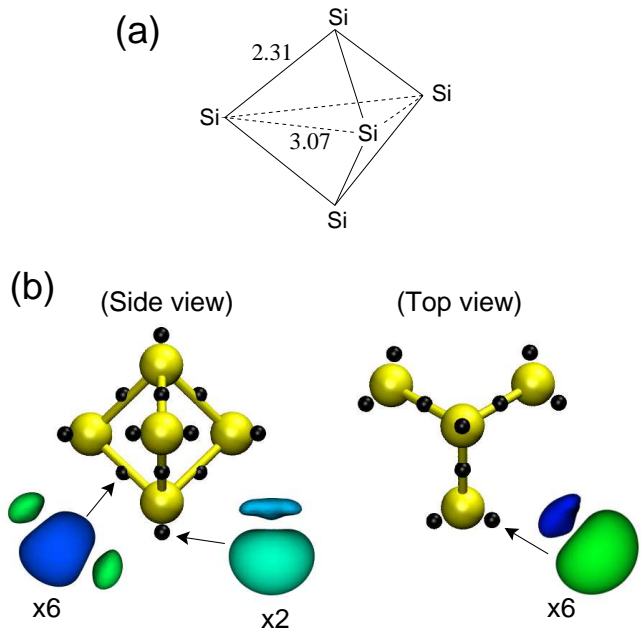}
\caption[cap.Pt6]{\label{fig.si5_orbitals} (a) Geometry of the
  $\text{Si}_5$ cluster. (b) Contour plots of the WFs corresponding to
  $L=4$. The position of the WF centers are indicated by black spheres.}
\end{figure}

\begin{figure}[!h]
\includegraphics[width=0.7\linewidth,angle=270]{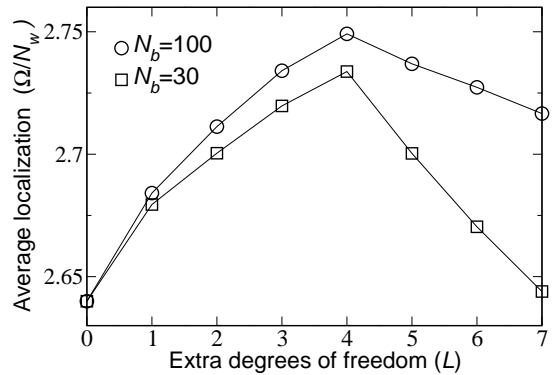}
\caption[cap.Pt6]{\label{fig.si5_spread} Average spread of the WFs of
the $\text{Si}_5$ cluster for different values of $N_b$ and $L$.}
\end{figure}

\end{subsection}

\begin{subsection}{Copper crystal}\label{sec.cufcc}
  To illustrate the method in the case of a periodic system we
  consider the construction of WFs for a copper crystal. This system
  was also studied by Souza \emph{et al.}\cite{souza01} using their
  disentangling method to obtain the WFs. Our results are in close agreement
  with those obtained by Souza \emph{et al.}, and this indicates
  the similarity of the two methods for periodic systems.

We use the primitive fcc unit cell and sample the first BZ on a uniform
(11,11,11) Monckhorst pack grid containing the $\Gamma$-point. To
obtain a minimal set of WFs describing the Cu $d$- and $s$-bands 
we set $N_w=6$. We construct two sets of WFs corresponding to
two different values of $E_0$: (i) $E_0=0.0$~eV
and (ii) $E_0=3.0$~eV, relative to the Fermi level. In the first case the
resulting WFs will span at least the occupied subspace and thus the
electronic structure described by the WFs will be correct below $E_F$. In
the second case the electronic structure will be correct up to 3~eV
above $E_F$, however, since this is a stronger restriction on the
localization space we must expect that the resulting WFs will be less localized
than those obtained in (i). In Figs.~\ref{fig.cuband1} and~\ref{fig.cuband2} we show
the original DFT bands together with the approximate bands computed by
diagonalizing the Hamiltonian within the subspace spanned by the 
WFs of case (i) and (ii), respectively. In both cases we see a very
good agreement between the exact and approximate bands below $E_0$. At
higher energies the approximate bands deviate from the exact bands,
indicating that the EDF which optimize the localization of the WFs do
\emph{not} coincide with specific Bloch eigenstates.
The quality of the WF bands below $E_0$ depends on the number of $\bold
k$-points used to construct the WFs. This is because the band diagram
must be constructed starting from fully localized functions, which
means that the coupling matrix elements must be truncated beyond a
cut-off distance given approximately by $N_i/2$ unit cells in the
direction $\bold a_i$. Thus the repeated cell, or equivalently the
number of $\bold k$-points, must be so large that the WFs have decayed
sufficiently between the repeated images.

\begin{figure}[!h]
\includegraphics[width=0.55\linewidth,angle=270,bb=127 1 556 745, clip]{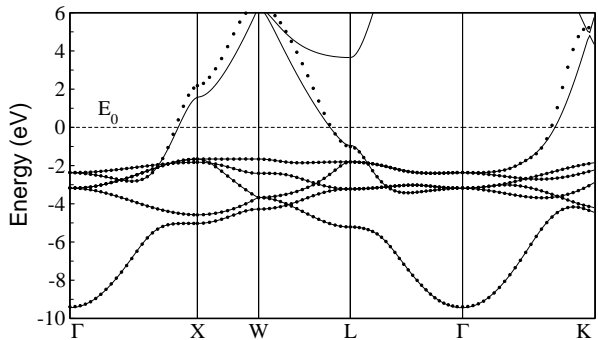}
\caption[cap.Pt6]{\label{fig.cuband1} Band structure of
  Cu(fcc). The full lines are the original DFT bands and the dots are
  the approximate bands computed from a set of six WFs ($N_w=6$). The WFs have been constructed using 11x11x11
  $\bold k$-points and keeping all occupied states in the localization
  space, i.e. $E_0=0.0$~eV relative to the Fermi level.}
\end{figure}

\begin{figure}[!h]
\includegraphics[width=0.55\linewidth,angle=270,bb=127 1 556 745, clip]{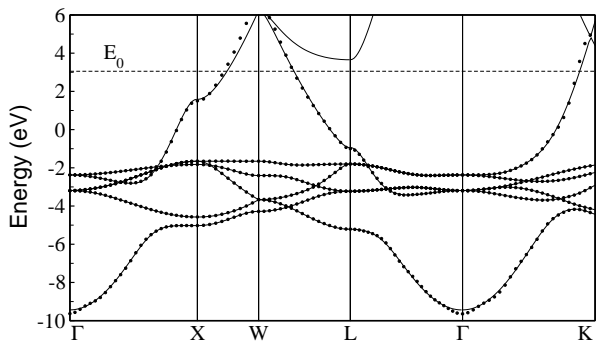}
\caption[cap.Pt6]{\label{fig.cuband2} Like Fig.~\ref{fig.cuband1}
  except that the WFs have been generated with $E_0=3.0$~eV.}
\end{figure}

Both sets of WFs consist of five atom-centered $d$-orbitals and a
single $s$-like orbital centered in one of the two interstitial sites.
The $d$-orbitals are more or less identical for the two cases, and
two examples are shown in Fig.~\ref{fig.cu_d}. Contour plots of the
$s$-like orbital is shown in Fig.~\ref{fig.cu_s} (b) and (c) for case (i)
and (ii), respectively. The plots indicate that the $s$-orbital of
case (ii) is less localized than the one obtained in case (i). That
this is indeed correct follows from the value of the spread
functional, $\Omega$, which is higher for (i) than for (ii).

\begin{figure}[!h]
\includegraphics[width=0.7\linewidth]{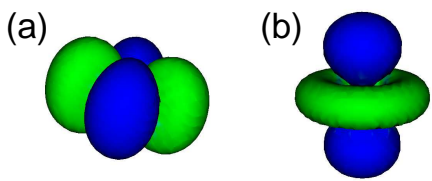}
\caption[cap.Pt6]{\label{fig.cu_d} Two $d$-like WFs for Cu(fcc). The
orbitals are centered at the atoms (not shown) .}
\end{figure}

\begin{figure}[!h]
\includegraphics[width=0.95\linewidth]{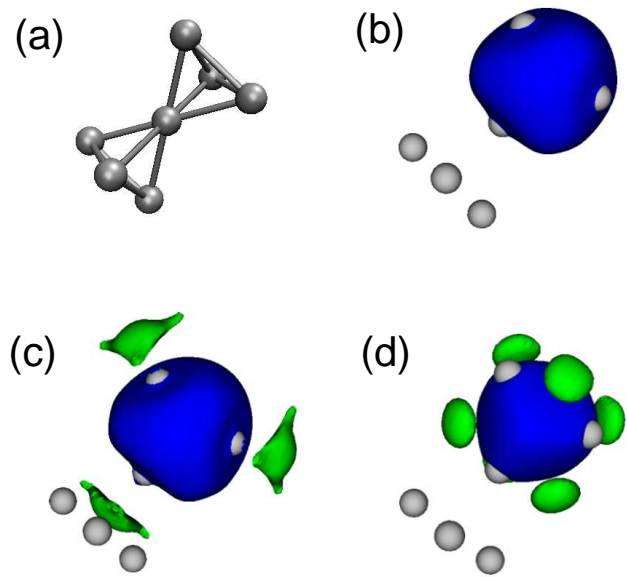}
\caption[cap.Pt6]{\label{fig.cu_s} (a) Two tetrahedral interstitial sites in the
  fcc crystal. (b-d) Contour plots of the s-like WF obtained for
  Cu(fcc). The WFs in (b) and (c) have been generated with $N_w=6$,
  $E_0=0.0$~eV and $N_w=6$, $E_0=3.0$~eV, respectively. Both WFs are
  located in one of the interstitial sites.  The WF in (d) correspond
  to $N_w=7$ and $E_0=0.0$~eV. In this case there is an equivalent WF
  located in the other interstitial site. The same contour value has been
  used for all plots.}
\end{figure}

\begin{figure}[!h]
\includegraphics[width=0.55\linewidth,angle=270,bb=127 1 556 745, clip]{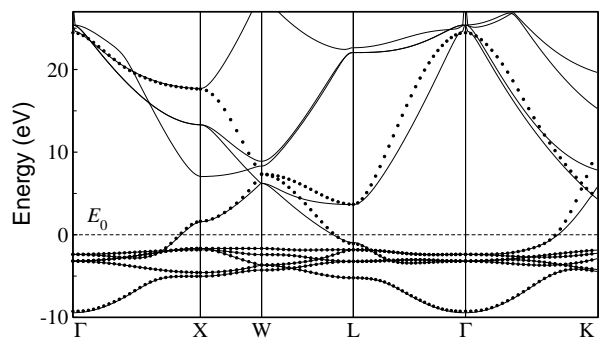}
\caption[cap.Pt6]{\label{fig.cuband3} Like Fig.~\ref{fig.cuband1}
  except that the WFs have been generated with $N_w=7$.}
\end{figure}

The minimal set of WFs obtained with $N_w=6$ breaks the symmetry of the fcc
crystal because the $s$-like orbital is located in one of the
interstitial sites leaving the other empty. As demonstrated by Souza
\emph{et al.} the symmetry can be restored by using seven
WFs per primitive cell instead of six. In Fig.~\ref{fig.cuband3} we
show the band structure obtained from a set of WFs
generated with $N_w=7$ and $E_0=0.0$~eV. We note that very high-energetic
states are now selected as the optimal EDF. This solution can
therefore only be obtained for rather large external localization
spaces, i.e. $N_b\geq 9$. The five $d$-like WFs are
unchanged, but now we obtain two equivalent $s$-like WFs located in
each of the two interstitial sites thereby restoring the fcc symmetry,
see Fig.~\ref{fig.cu_s}(d). We have calculated the average
localization $\langle \Omega \rangle$ for $N_w=6,7,8$, and found that
the maximum is attained for the symmetric solution with $N_w=7$.

\end{subsection}

\begin{subsection}{Nitrogen absorption on Cu(100)}\label{sec.ncu}
  In this section we study the WFs of a copper (100)-surface covered
  with half a mono-layer of nitrogen atoms. As the system is neither periodic (in
  all directions) nor isolated, it represents a very general
  situation. The section is divided into two parts. In the first part
  the WFs are constructed and analyzed, and in the second part we use the
  obtained WFs to study the chemisorption of nitrogen within the Newns
  Anderson model.
  
\begin{subsubsection}{Wannier function analysis}
  We model the Cu(100) surface by a slab with a thickness of two
  atomic layers. The supercell contains four Cu atoms and a single N atom
  adsorbed in a hollow site, and its height is such that the surface slabs are
  separated by 9.0~\AA~of vacuum. A topographic top-view of the
  surface is shown in Fig.~\ref{fig.NCu_topograph}. We sample the
  first BZ
  on a uniform (7,7,1) Monckhorst pack grid containing the
  $\Gamma$-point.

\begin{figure}[!h]
\includegraphics[width=0.67\linewidth]{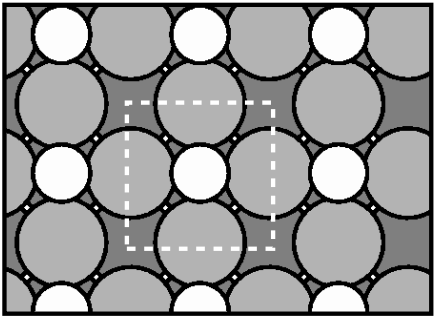}
\caption[cap.Pt6]{\label{fig.NCu_topograph} Topographic view on the
  Cu(100) surface with adsorbed nitrogen. The white spheres are
  nitrogen, while the light gray spheres represent the Cu surface
  layer. A supercell is indicated.}
\end{figure}
\begin{figure}[!h]
\includegraphics[width=0.82\linewidth,angle=270]{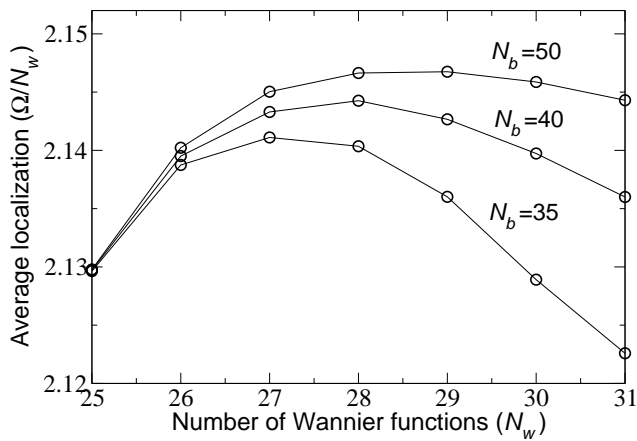}
\caption[cap.Pt6]{\label{fig.NCu_totalspread} Average localization of
  the WFs of the nitrogen covered Cu surface for different values of
  $N_b$ and $N_w$.}
\end{figure}

Let us start by considering what we can expect to find on the
basis of our previous experience. First, the result from the copper
crystal suggests that a minimal description of the metal
surface is obtained with five $d$-orbitals and an $s$-like orbital per
Cu atom. Since there are four Cu atoms per supercell this gives a
total of 24 WFs.  Next, the similarity between the valency of N and Si
together with our experience from the $\text{Si}_5$ cluster points to a description of
the nitrogen atom in terms of $sp^x$-hybrides.

In Fig.~\ref{fig.NCu_totalspread} we have plotted the average
localization, $\langle \Omega \rangle$, of the obtained WFs as a function of $N_w$ for
three different sizes of the external localization space corresponding to
$N_b=35,40,50$. In all cases we have set $E_0=E_F$ in order to ensure
that the occupied eigenstates are exactly reproduced by the WFs. As
expected, the localization improves as the size of the external
localization space increases. In addition, the maximum of $\langle \Omega \rangle$ shifts towards
larger $L$-values as $N_b$ is increased. Specifically the maximum 
shifts from $N_w=27$ to $N_w=29$ as $N_b$ is increased from 35 to 50. This
is not unexpected since we know that $\langle \Omega \rangle$ will be a monotonically
increasing function of $L$ in the limit $N_b\to \infty$, see discussion
in Sec.~\ref{sec.optimal}. Again we stress that it is only the degree
of localization of the WFs that change with $N_b$ for a fixed $L$, and
not their qualitative form. Thus the chemical picture provided by the
WFs does not change with $N_b$. In fact, for all the values of $N_b$
we obtain 20 highly localized $d$-orbitals (five located on each of
the four Cu atoms) and four $sp^3$ orbitals centered on the N atom,
see Fig.~\ref{fig.NCu_orbitals}. The
remaining $N_w-24$ WFs are the less localized $s$-like orbitals of Cu.
Thus, as $N_w$ is increased beyond 24, the number of $s$-like Cu WFs
simply increases correspondingly.

\begin{figure}[!h]
\includegraphics[width=0.82\linewidth,angle=270]{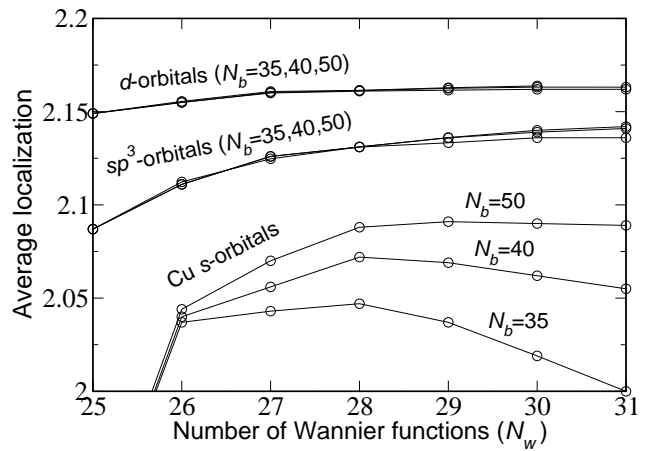}
\caption[cap.Pt6]{\label{fig.NCu_spread} Average localization of
  the $d$-, $sp^3$- and $s$-like WFs considered separately for
  different values of $N_b$ and $N_w$.}
\end{figure}

To gain further insight into the dependence of the WFs on
$N_b$ and $N_w$, we show in Fig.~\ref{fig.NCu_spread} the average localization of
the $d$-, $sp^3$- and $s$-orbitals, separately.  It is clear that the
$N_b$-dependence as well as the maximum of $\langle \Omega \rangle$ are almost exclusively
related to the Cu $s$-orbitals.  Except for the case $N_b=50$, which
is in fact somewhat extreme since states of 20 eV above the Fermi
level are included in the external localization space, the average
spread of the Cu $s$-orbitals is maximal for $N_w=28$. This
corresponds to one $s$-orbital per Cu which is exactly what we
anticipated from the analysis of the copper crystal.

We end by summarizing the chemical picture obtained from the WF analysis: For
$N_w=28$ the Cu surface is described by the minimal set of WFs
consisting of five $d$- and one $s$-like orbital per atom. For the
nitrogen we obtain four $sp^3$ hybrids oriented as indicated
in Fig.~\ref{fig.NCu_orbitals}.

\begin{figure}[!h]
\includegraphics[width=0.9\linewidth]{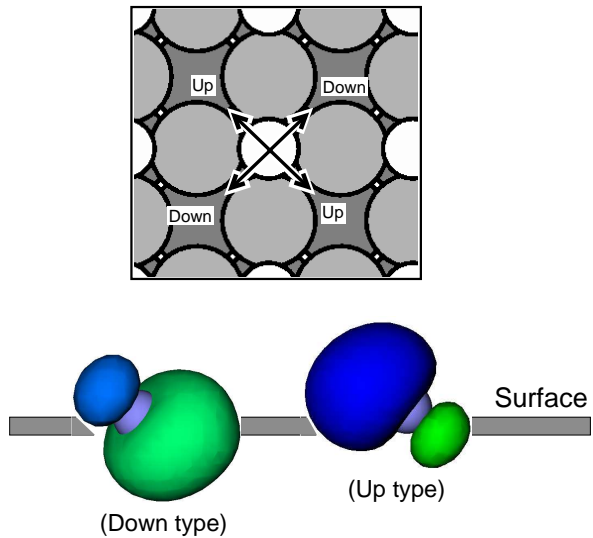}
\caption[cap.Pt6]{\label{fig.NCu_orbitals} 
  Orientation of the four $sp^3$-like WFs belonging to the N
  atom. Seen from above the orbitals point to the bridge sites of the
  Cu atoms in the surface. Each orbital points either up from
  or down into the surface. An example of each type is shown in the
  lower panel.}
\end{figure}

\end{subsubsection}

\begin{subsubsection}{Adsorption in the Newns Anderson model}
  The WFs can be used to obtain a detailed and consistent picture of
  the hybridization occurring between the nitrogen states and the
  states of the substrate.  As we shall see the analysis gives a
  complete account for the shape of the projected density of states of
  a
  given N orbital, in terms of the bare orbital energy, a coupling
  strength, and the density of states of the so-called group orbital.

In the Newns Anderson model, one considers an adsorbate 
state, $|a\rangle$, of energy $\varepsilon_a=\langle a|H|a\rangle$, coupled to a continuum of states,
$|k\rangle$, representing the substrate. The
coupling matrix elements are denoted by $V_k=\langle a|H|k
\rangle$. A particularly useful formulation can be obtained by
introducing the normalized group orbital, $|g\rangle=V^{-1}\sum_k V_{k}|k\rangle$, where $V=(\sum_k
|V_k|^2)^{-1/2}$. It is easily checked, that the coupling between
$|a\rangle$ and any substrate state orthogonal to $|g\rangle$
vanishes. Consequently $|a\rangle$ is coupled to the substrate via the
group orbital only, and the coupling is given by $V$, i.e. $V=\langle
a|H|g \rangle$.

Physical quantities such as the projected density of states (PDOS) of
the adsorbate state and the hybridization part of the adsorption energy,
can be obtained from the retarded adsorbate Green's function, which in
turn follows from the three quantities $\varepsilon_a$, $V$ and 
$\rho^0_g(\varepsilon)$, where $\rho^0_g$ denotes the PDOS of the
group orbital in the absence of coupling to the adsorbate state. Often
$\rho^0_g(\varepsilon)$ is referred to as the band to which the
adsorbate is coupled.

\begin{figure}[!h]
\includegraphics[width=0.85\linewidth,angle=270]{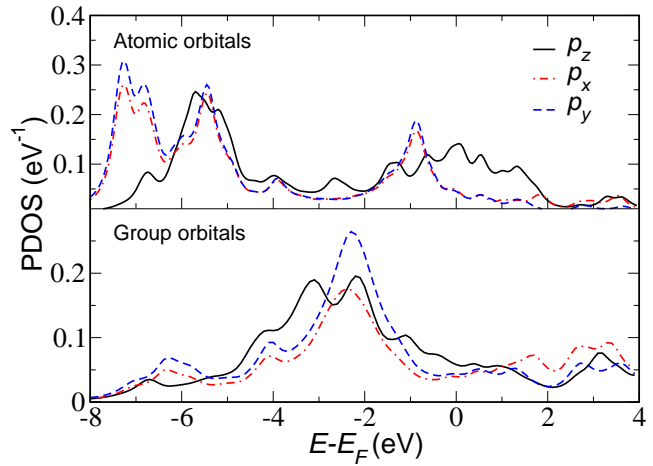}
\caption[cap.Pt6]{\label{fig.NCu_LDOS} Top: PDOS for the atomic
  $p$-orbitals obtained by diagonalizing the Hamiltonian in the
  subspace spanned by the $sp^3$ WFs of the N atom. Bottom: PDOS for
  the group orbitals corresponding to each of the atomic
  $p$-orbitals.}
\end{figure}

The $sp^3$ WFs of the N atom are not well suited as a starting point
for applications of the Newns Anderson model, since they do not
represent the energy levels of the free atom. This problem can be
overcome by diagonalizing the Hamiltonian matrix in the WF basis,
within the subspace spanned by the four $sp^3$ orbitals. The result of
the subspace diagonalization is a set of four atomic orbitals
consisting of one $s$-like and three $p$-like orbitals, each centered
at the N atom. Two of the $p$-orbitals lie in the surface plane (the
$xy$-plane) and are directed along the arrows shown in
Fig.~\ref{fig.NCu_orbitals}, while the third is oriented along the
surface normal (the $z$-axis). We shall refer to the $p$-orbitals as
$p_x, p_y$ and $p_z$, respectively. The energies corresponding to the
atomic orbitals are (in eV): $\varepsilon_s=-14.8$,
$\varepsilon_z=-2.4$, $\varepsilon_x=-3.7$, and $\varepsilon_y=-4.2$.
We notice, that the energy of the $p_x$ and $p_y$ orbitals differ even
though the symmetry of the system suggests that they should be equal.
The reason for this is that the WFs break the four-fold rotation
symmetry of the system, i.e. the subspace spanned by the four $sp^3$
WFs is not invariant under the same symmetry transformations as the
Hamiltonian. This is not surprising, since the WFs are constructed
solely from a criterion of maximal localization and no attempts are
made to conserve symmetries. On the other hand we have found that by
increasing the parameter $E_0$ above the value $E_0=E_F$ used in the
present example, the symmetry between $p_x$ and $p_y$ can be restored. The
price one has to pay is that the copper $s$-like WFs become less
localized due to the further constrains on the localization space
implied by the larger value of $E_0$. From the Hamiltonian in the WF basis we
can also obtain the coupling, $V$, between each of the atomic nitrogen
orbitals and its corresponding group orbital. These are quite similar
and vary from $3.1$~eV to $3.8$~eV.

In Fig.~\ref{fig.NCu_LDOS} we show the calculated PDOS for each of the
three nitrogen $p$-orbitals (upper panel). Although the on-site
energies of the $p_x$ and $p_y$ orbitals differ (as discussed above),
their PDOS are rather similar. The PDOS of the corresponding group
orbitals have been calculated with all coupling matrix elements to the
N orbitals set to zero, i.e. the adsorbate states have effectively
been decoupled from the surface. The result is shown in the lower
panel of Fig.~\ref{fig.NCu_LDOS}. For all three orbitals, the on-site
energies lie within the band. Due to the strong coupling, bonding and
anti-bonding resonances are formed at the band edges around $-7$~eV
and $0$~eV as can be seen in the upper panel of the figure.  This is
the limit of strong chemisorption.~\cite{newns69} Since the four
orbitals span all states with significant weight on the N atom, this
representation provides a full representation of the nitrogen bonding.

\end{subsubsection}
\end{subsection}
\end{section}

\begin{section}{Conclusions}
  We have presented a practical method for constructing partly
  occupied WFs for a wide range of systems. The method employs a bonding-antibonding closing
  procedure to filter out a set of unoccupied states, called the extra
  degrees of freedom, which serve to improve the localization of the
  WFs. The determination of the extra degrees of freedom is based on a
  minimization of the spread of the resulting WFs. We derived
  expressions for the gradients of the spread functional and showed
  how these can be combined with a Lagrange multiplier scheme to
  minimize the spread functional.
  
  The generality of the scheme was demonstrated by applying the method to
  three different systems. As an example of an isolated system, we
  considered a $\text{Si}_5$ cluster, and showed
  how different sets of WFs could be obtained by varying the number of
  extra degrees of freedom. A similar analysis was performed for a
  copper crystal, where we found results very similar to those of
  Souza \emph{et al.}~\cite{souza01}. Finally we studied in detail the
  WFs of a Cu(100) surface with a nitrogen coverage of 0.5. In many
  cases we were able to obtain a special set of WFs with a
  particularly high degree of symmetry and localization, by maximizing
  the average spread of the WFs. Moreover, the condition of maximal average
  localization was shown to coincide with a complete matching of
  bonding and antibonding states.

\end{section}

\begin{section}{Acknowledgments}
We acknowledge support
from the Danish Center for Scientific Computing through Grant No.
HDW-1101-05.
\end{section}

\appendix
\begin{section}{Spread functional for Vanderbilt ultrasoft pseudo-potentials}

For Vanderbilt ultra-soft pseudo-potentials\cite{vanderbilt90} the
optimal smoothness of the pseudo-wavefunctions is obtained by relaxing
the norm-conserving constrains for the pseudo-wavefunctions. This
results in a generalized orthonormality relation\cite{vanderbilt90}
\begin{equation}\label{eq.gen_orth}
   \langle \psi_i | S | \psi_j \rangle = \delta_{ij}.
\end{equation}
The Hermitian operator $S$ is given by
\begin{equation}
   S = 1 + \sum_I \sum_{nm} q_{nm} |\beta_n^{I}\rangle\langle{\beta_m}^{I}|,
\end{equation}
where the index $I$ denotes the atoms in the system, and $q_{nm}$ is given by
\begin{equation}
q_{nm}=\int{d{\bold r}Q_{nm}^{I}(\bold r)}.
\end{equation}
The functions $\{\beta_n^{I}\}$ and $\{Q_{nm}^{I}\}$ are all localized functions 
centered at atom $I$. The functions $\{Q_{nm}^{I}\}$ describe the
augmentation charge not contained in the smooth pseudo-wavefunctions, and
they must therefore be included in the calculation of the spread of the wavefunctions.

\begin{subsection}{Large supercells}
\label{vanderbilt.gamma}

In the case of large supercells, using the $\Gamma$-point approximation, 
Bernasconi and Madden\cite{bernasconi_madden01} derived the following expression for 
the contribution to $Z^{(0)}_{\alpha}$ from the augmentation
charges $Q_{nm}^{I}(\bold r)$:

\begin{equation}\label{eq.zus_gamma}
 Z^{(us,0)}_{\alpha,ij} = \sum_{I,nm} \langle \psi_{i} | \beta_m^{I}\rangle \langle \beta_{n}^I | \psi_j\rangle
                          \int d\bold r e^{-\text{i} \bold G_{\alpha}\cdot \bold r} Q_{mn}(\bold r)
\end{equation}

\end{subsection}

\begin{subsection}{Periodic systems}
\label{vanderbilt.kpoints}
For the periodic case, using a uniform $\bold k$-point grid, we write $Z_{\alpha}^{us}$ as
\begin{equation}
Z_{\alpha}^{us} = \sum_{\bold k\bold k'} Z_{\alpha}^{(us)\bold k\bold{k'}}.
\end{equation}
Here again we use the exact correspondence between the Bloch states ${\psi_{n\bold k}}$ and the $\Gamma$-point
eigenstates of the repeated cell.
In the repeated cell we use the notation,
\begin{equation}\label{eq.beta}
  h_{i\bold k}^{In\bold{t}} = \langle \psi_{i\bold k} | \beta_n^{I,\bold t}\rangle =  
                               \langle \psi_{i\bold k} |
                               \beta_n^{I,\bold t=\bold{0}}\rangle
                               e^{i\bold k \cdot \bold R_{\bold t}}
\end{equation}
and
\begin{equation}\label{eq.qnmexp}
Q_{nm}^{I \bold t} = \sum_{\bold{G}} Q_{nm}(\bold{G})
e^{-i\bold{G}\cdot (\bold r-\bold{R_{\bold t}})}
\end{equation}
$\bold R_{\bold{t}}$ is here a real space translation vector, given in terms of the basis
$\bold a$, $t_1\bold a_1 + t_2\bold a_2 +t_3\bold a_3$, $\bold t = (t_1,t_2,t_3)$. 
We will use $h^{In} = h^{I,\bold t=\bold{0},n}$ in what follows. 
Inserting $h_{i\bold k}^{In\bold t}$ and $Q_{nm}^{I\bold t}$ from Eqs.~(\ref{eq.beta}) and ~(\ref{eq.qnmexp}), 
together with the Bloch states, ${\psi_{i\bold k}}$, into the
$\Gamma$-point expression for $Z_{\alpha}^{(us,0)}$ in Eq.~(\ref{eq.zus_gamma}), we find
\begin{equation}
  Z_{\alpha,ij}^{(us,0),\bold k\bold k'} =  \sum_{\bold{t}} \sum_{I,nm} h_{i\bold k}^{In\bold t} h_{j\bold k'}^{Im\bold t}
                \int \text{d}\bold r e^{-i\bold G_{\alpha} \bold{r}}
                Q_{nm}^{I \bold t}(\bold r)
\end{equation}
The sum over $\bold t$ is for $t_i=0,..,N_i-1$, see Eq.~(\ref{eq.kpoints}). 
Inserting the left hand side of Eqs.~(\ref{eq.qnmexp}) and ~(\ref{eq.beta}), and rearranging, we find 
\begin{widetext}
\begin{equation}
Z_{\alpha,ij}^{(us,0),\bold k\bold k'} =
  \sum_{I,nm} h_{i\bold k}^{In} h_{j\bold k'}^{Im}
  \sum_{\bold{t}} e^{-i(\bold k -\bold k')\cdot \bold R_{\bold{t}}}
  \sum_{\bold{G}} Q_{nm}(\bold G) e^{-i\bold G \cdot \bold R_{\bold{t}}} \int{ \text{d}\bold r e^{i(\bold G - \bold G_{\alpha})\cdot \bold r}}.
\end{equation}
Finally, we arrive at our expression for $Z_{\alpha,ij}^{(us,0),\bold k\bold k'}$
\begin{equation}
 Z_{\alpha,ij}^{(us,0),\bold k\bold k'} =
   \sum_{I,nm} h_{i\bold k}^{In} h_{j\bold k'}^{Im}
   \sum_{\bold t} e^{-i (\bold k -\bold{k'} - \bold G_{\alpha})\cdot \bold R_{\bold{t}}} Q_{nm}(\bold G=\bold G_{\alpha}),
\end{equation}

\end{widetext}
which is non-zero only when $\bold k$ and $\bold k'$ fulfills the condition in Eq.~(\ref{eq.condition}).
Again we see that this expression contains the $\Gamma$-point formalism, Eq.~(\ref{eq.zus_gamma}),
as a special case.

\end{subsection}

\end{section}
\newpage


\bibliographystyle{apsrev}

\end{document}